\documentclass[aps, prl, twocolumn, amsmath, amssymb, superscriptaddress, nofootinbib]{revtex4-1}

\usepackage[utf8]{inputenc}
\usepackage{graphicx}
\usepackage{units}
\usepackage{color}
\usepackage[pdftex, colorlinks=true, linkcolor=myblue, citecolor=myblue, urlcolor=myblue]{hyperref}
\usepackage{times}
\usepackage{comment}
\usepackage{graphicx}
\usepackage{amsmath, calc}
\usepackage{mathrsfs}
\usepackage{amsfonts}
\usepackage{amssymb}
\usepackage{bm}
\usepackage{bbm}
\usepackage{color}
\usepackage{array}
\usepackage{units}

\definecolor{myblue}{rgb}{0,0,1}
\definecolor{myred}{rgb}{1,0,0}

\begin{document}
%\narrowtext
\title{Quasi-Chiral Interactions between Quantum Emitters at the Nanoscale}

\author{C.~A.~Downing}
\affiliation{Departamento de F\'{i}sica T\'{e}orica de la Materia Condensada and Condensed Matter Physics Center (IFIMAC),
Universidad Aut\'{o}noma de Madrid, E-28049 Madrid, Spain}

\author{J.~C.~L\'{o}pez~Carre\~{n}o}
\affiliation{Faculty of Science and Engineering, University of Wolverhampton, Wulfruna Street, Wolverhampton WV1 1LY, UK}
\affiliation{Departamento de F\'{i}sica T\'{e}orica de la Materia Condensada and Condensed Matter Physics Center (IFIMAC),
Universidad Aut\'{o}noma de Madrid, E-28049 Madrid, Spain}

\author{F.~P.~Laussy}
\affiliation{Faculty of Science and Engineering, University of Wolverhampton, Wulfruna Street, Wolverhampton WV1 1LY, UK}
\affiliation{Russian Quantum Center, Novaya 100, 143025 Skolkovo, Moscow Region, Russia.}

\author{E.~del Valle}
\email{elena.delvalle@uam.es}
\affiliation{Departamento de F\'{i}sica T\'{e}orica de la Materia Condensada and Condensed Matter Physics Center (IFIMAC),
Universidad Aut\'{o}noma de Madrid, E-28049 Madrid, Spain}

\author{A.~I.~Fern\'{a}ndez-Dom\'{i}nguez}
\email{a.fernandez-dominguez@uam.es}
\affiliation{Departamento de F\'{i}sica T\'{e}orica de la Materia Condensada and Condensed Matter Physics Center (IFIMAC),
Universidad Aut\'{o}noma de Madrid, E-28049 Madrid, Spain}

\date{\today}

%===========================================================================
%===========================================================================
%===========================================================================

\begin{abstract}
We present a combined classical and quantum electrodynamics
description of the coupling between two circularly-polarized
quantum emitters held above a metal surface supporting surface
plasmons. Depending on their position and their natural frequency,
the emitter-emitter interactions evolve from being reciprocal to
non-reciprocal, which makes the system a highly tunable platform
for chiral coupling at the nanoscale. By relaxing the stringent
material and geometrical constraints for chirality, we explore the
interplay between coherent and dissipative coupling mechanisms in
the system. Thus, we reveal a quasi-chiral regime in which its
quantum optical properties are governed by its subradiant state,
giving rise to extremely sharp spectral features and strong photon
correlations.
\end{abstract}

%===========================================================================
%===========================================================================
%===========================================================================

\maketitle

A quarter of a century ago, the theory of cascaded quantum systems was
developed independently by Gardiner and Carmichael~\cite{Gardiner1993,
  Carmichael1993, Gardiner2004}. By construction, the theory describes
distant source-target systems whereby non-reciprocal, unidirectional
interactions arise naturally: the former is coupled to the latter
while completely forbidding the opposite. Today, the emerging field of
chiral quantum optics~\cite{Lodahl2017} seeks to realize and exploit systems
exhibiting non-reciprocal light-matter
interactions. Properly harnessed, the chiral coupling between quantum
emitters (QEs) and photons at the quantum level promises a myriad of
nontrivial applications in quantum communication and information, including non-reciprocal single-photon
devices~\cite{Shomroni2014}, optical isolators~\cite{Sayrin2015},
optical circulators~\cite{Sollner2015} and integrated quantum
optical circuits~\cite{Barik2018}.

In this Letter, we present a classical and quantum electrodynamics
description of the most elemental physical platform yielding
non-reciprocal interactions at the nanoscale: two
circularly-polarized QEs placed on top of a flat metal surface
supporting tightly confined surface plasmons (SPs). The high
tunability of the system, which can be manipulated through the
relative position of the QEs, their natural frequencies, and the
distance from the metal surface, enables us to unveil a rich
landscape of coherent and dissipative emitter-emitter couplings.
This includes the much sought-after chiral (fully non-reciprocal)
regime~\cite{Lodahl2017}. Through a comparison with the
cascaded formalism~\cite{Metelmann2015}, we set a well-defined
criterion for chiral coupling. Moreover, by relaxing the geometric
and material parameters satisfying the unidirectional conditions,
we investigate the evolution of the one- and two-photon far-field
spectra of the system in the transition from the conventional
(reciprocal) regime. This allows us to identify a less stringent,
more easily accessible quasi-chiral configuration, in which the
interaction between QEs is highly directional, and where the system
develops extremely sharp spectral features and strong photon
correlations. Our approach does not break Lorentz reciprocity \cite{Caloz2018} since
the electromagnetic (EM) fields maintain time-reversal symmetry,
and it is only the emitter-emitter interactions mediated
by them that become unidirectional \cite{Fan2012}.

%===========================================================================
%===========================================================================
%===========================================================================

The starting point of our description for the two QEs and their
EM coupling is the Hamiltonian $H = \omega_0 (
\sigma_1^{\dagger} \sigma_1 + \sigma_2^{\dagger} \sigma_2 ) +
g_{12} \sigma_1^{\dagger} \sigma_2 + g_{21} \sigma_2^{\dagger}
\sigma_1$, where $\sigma_{i}$ ($\sigma_{i}^{\dagger}$) is the
lowering (raising) operators of QE-$i$, and $\omega_0$ is its
natural frequency. Hermiticity dictates that $g_{12} =
g_{21}^{\ast}$, which makes coherent interactions fully
reciprocal. We introduce damping in our model in the form of
the inherent decay of both QEs and their dissipative coupling
(collective decay). Both mechanisms are accounted for by the
Master equation~\cite{delValle2010, MartinCano2011} $\partial_t
\rho = \mathrm{i} [ \rho, H ] + \sum_{i, j = 1, 2} (\gamma_{i
j}/2) \mathscr{L}_{ij} \rho + \sum_{i = 1, 2} (P_{i}/2) \left(
\mathscr{L}_{ii} \rho \right)^\dagger$, where $\gamma_{i j}$ stand
for self ($i=j$) and collective ($i \ne j$) decay rates, and
$\mathscr{L}_{ij} \rho = 2 \sigma_j \rho \sigma_i^{\dagger} -
\sigma_i^{\dagger} \sigma_j \rho - \rho \sigma_i^{\dagger}
\sigma_j$ is the Lindblad superoperator ($\hbar = 1$). Note
that we have also included the incoherent driving rates $P_{i}$,
which feed population into both QEs (within the vanishing pump
limit)~\cite{supp}. Similar to their coherent counterparts, the collective
dissipative constants satisfy $\gamma_{12} =
\gamma_{21}^{\ast}$.

The Master equation above encompasses a rich
phenomenology as a function of the coupling parameters $\{ g_{12},
\gamma_{12} \}$, which are in general complex quantities.
Interestingly, it maps onto the cascaded Master equation if
\begin{equation}
\frac{|g_{12}|}{|\gamma_{12}|} = \frac{1}{2} , \quad
\mathrm{arg}\Big(\frac{g_{12}}{\gamma_{12}}\Big) = \frac{3\pi}{2}.
\label{eq:conditions}
\end{equation}
The two equations above establish the magnitude and phase balance in
the coherent and dissipative components of the QE interactions that
give rise to chiral coupling. Essentially, Eqs.~\eqref{eq:conditions}
describe the effective isolation of QE-2 from the output of QE-1. Note that
the damping rates must satisfy $0 \le | \gamma_{12} | \le \gamma_{11}$
(from now on, we assume $\gamma_{22}=\gamma_{11}$), and complete
equivalence to the cascaded formalism occurs only in the limit
$|\gamma_{12}|=\gamma_{11}=\gamma_{22}$~\cite{Metelmann2015}.

\begin{figure}[!ht]
 \includegraphics[width=0.95\linewidth]{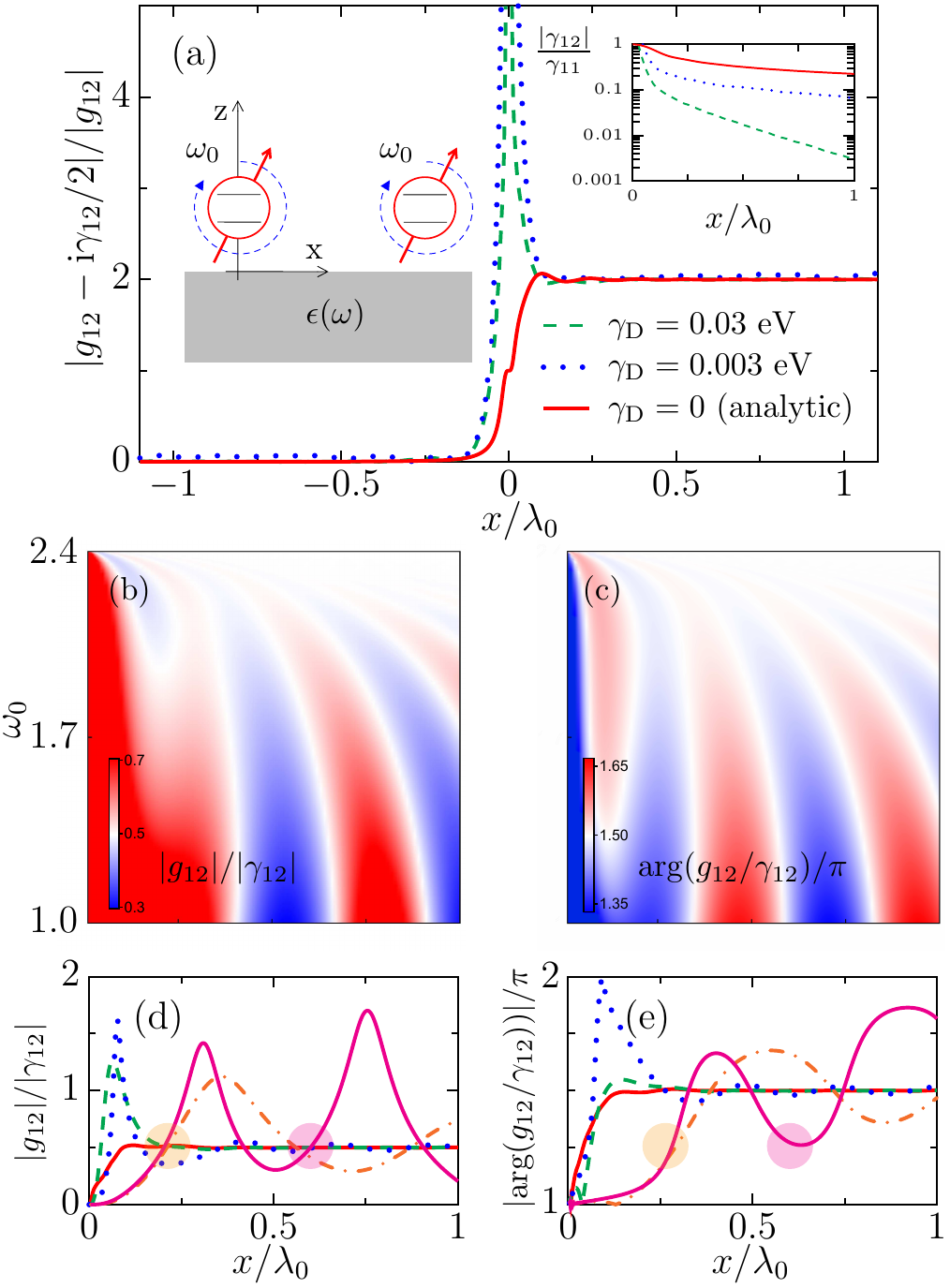}
  \caption{(a) Effective coupling between
two circularly-polarized QEs above a metal surface (left inset)
versus $x/\lambda_0$ ($z=5$~nm, $\omega_0=2.35$ eV). Right inset:
dissipative coupling versus $x/\lambda_0$. Relative magnitude (b)
and phase (c) of coherent and dissipative couplings versus
$\omega_0$ and $x/\lambda_0$. Solid red lines in (d) and (e) are
cuts of panels (b) and (c) at $2.35~\mathrm{eV}$. Plots for four
larger $z$-values and $\gamma_{\rm D}=0.03$ eV are shown: 30
(green dashed), 60 (blue dotted), 150 (orange dot-dashed) and 300
nm (solid pink). Shaded circles indicate configurations satisfying
$|g_{12}|/|\gamma_{12}| = 1/2$ and
$\mathrm{arg}(g_{12}/\gamma_{12}) = 5\pi/4$ for the last two
$z$-values.}
 \label{fig:fig1}
\end{figure}

We parameterize our Master equation using EM calculations for the
setup sketched in the left inset of Fig.~\ref{fig:fig1}~(a), which
mimics recent setups realizing unidirectional plasmon emission in
metallic systems~\cite{Rodriguez2013,OConnor2014}. We model the
metal response through a Drude permittivity
$\epsilon(\omega)=1-\omega_{\mathrm{pl}}^2/\omega(\omega+\mathrm{i}\gamma_{\mathrm{D}})$,
with plasma frequency $\omega_{\mathrm{pl}} = 3.4~\mathrm{eV}$
(which corresponds to an asymptotic SP frequency
$\omega_{\mathrm{sp}}\simeq \omega_{\mathrm{pl}} /
\sqrt{2}=2.40~\mathrm{eV}$) and investigate the impact of metal
absorption by varying the Drude damping $\gamma_{\mathrm{D}}$. The
dipole moment of the QEs is ${\mathbf{d} = |\mathbf{d}| \left(
\hat{x} + \mathrm{i} \hat{z}
  \right) / \sqrt{2}}$, whose circularly-polarization opens up
novel degrees of freedom via the phases of the coherent and
dissipative coupling parameters. Within the formalism of macroscopic
quantum electrodynamics~\cite{Dung2002}, the coupling parameters have
the form ${g_{ij} = \omega_0^2 \mathbf{d}_i^\ast \mathrm{Re} \left\{
  \mathbf{G} \left( \mathbf{r}_i, \mathbf{r}_j, \omega_0 \right)
  \right\} \mathbf{d}_j /\epsilon_0 c^2}$ and ${\gamma_{ij} = 2
  \omega_0^2 \mathbf{d}_i^\ast \mathrm{Im} \left\{ \mathbf{G} \left(
  \mathbf{r}_i, \mathbf{r}_j, \omega_0 \right) \right\}
  \mathbf{d}_j/\epsilon_0 c^2}$, where $\mathbf{G}(\cdot)$ stands for
the EM dyadic Green's function (DGF).

To gain insight into the emergence of chirality in our system, we
compare numerical solutions of Maxwell's equations against
analytical predictions for the coupling strengths obtained by
keeping only the plasmon-pole contribution to the DGF~\cite{Nikitin2009,supp}.
From now on, we assume that both
QEs are placed at the same height $z$, and introduce the position of QE-2
relative to QE-1, $x=x_2-x_1$. In the limit of
vanishing metal losses and for $k_{\rm sp}(\omega_0)|x|\gg 1$
(where $k_{\rm
sp}(\omega_0)=(\omega_0/c)\epsilon(\omega_0)/\sqrt{\epsilon(\omega_0)+1}$
is the SP wavevector at the QE frequency), they read
\begin{equation}
g_{12} = \eta(\omega_0)\frac{e^{-2\sqrt{k_{\rm
sp}(\omega_0)^2-(\omega_0/c)^2}z}}{\sqrt{2\pi \mathrm{i} k_{\rm
sp}(\omega_0)|x|}}e^{\mathrm{i}k_{\rm
sp}(\omega_0)x}=g_{21}^\ast,\label{eq:g12}\\
\end{equation}
and
\begin{equation}
\gamma_{12} = 2 \mathrm{i}\,{\rm
sgn}(x)\,g_{12}=\gamma_{21}^\ast,\label{eq:gamma12}
\end{equation}
with
$\eta(\omega)=(\omega/c)^3|\mathbf{d}|^2\epsilon(\omega)^2/\epsilon_0[\epsilon(\omega)+1]^{5/2}[\epsilon(\omega)-1]$.
Importantly, Eqs.~\eqref{eq:g12}~and~\eqref{eq:gamma12} naturally
satisfy Eqs.~\eqref{eq:conditions}, which proves that the QEs become
chirally coupled when their interaction is fully mediated by confined
SPs~\cite{Lodahl2017}.

Figure~\ref{fig:fig1}(a) plots the absolute value of the effective
coupling strength $|g_{12} - \mathrm{i} \gamma_{12}/2|$, which
combines the dissipative and coherent components, normalized to
$|g_{12}|$, as a function of the distance between the two QEs
($\omega_0=2.35$ eV, $\lambda_0=527$ nm, and $z=5$ nm). Numerical
calculations for two different $\gamma_{\rm D}$ (green dashed and
blue dotted lines) are compared against the lossless analytical
result (red solid line). Remarkably, for $|x|\gtrsim
0.1\,\lambda_0\simeq 53$ nm, the QE interactions are governed by
SPs (analytical and numerical predictions are in perfect
agreement) and the effective coupling becomes directional: its
magnitude approaches $2|g_{12}|$ for ${\rm
  sgn}(x)>0$ and vanishes for ${\rm sgn}(x)<0$. We also observe
that the inclusion of metal absorption does not degrade the chiral
character of the QE coupling, which is insensitive to $\gamma_{\rm
  D}$. The right inset shows that the only effect of the Drude losses
is to reduce the efficiency of the dissipative coupling mechanism,
weighted by $|\gamma_{12}|/\gamma_{11}$.  Notably, recent reports
have demonstrated that this ratio (termed the $\beta$-factor)
can be optimized in complex plasmonic
structures~\cite{Akimov2007,Bermudez2015} beyond our
proof-of-principle proposal.

\begin{figure*}[t]
 \includegraphics[width=0.95\linewidth]{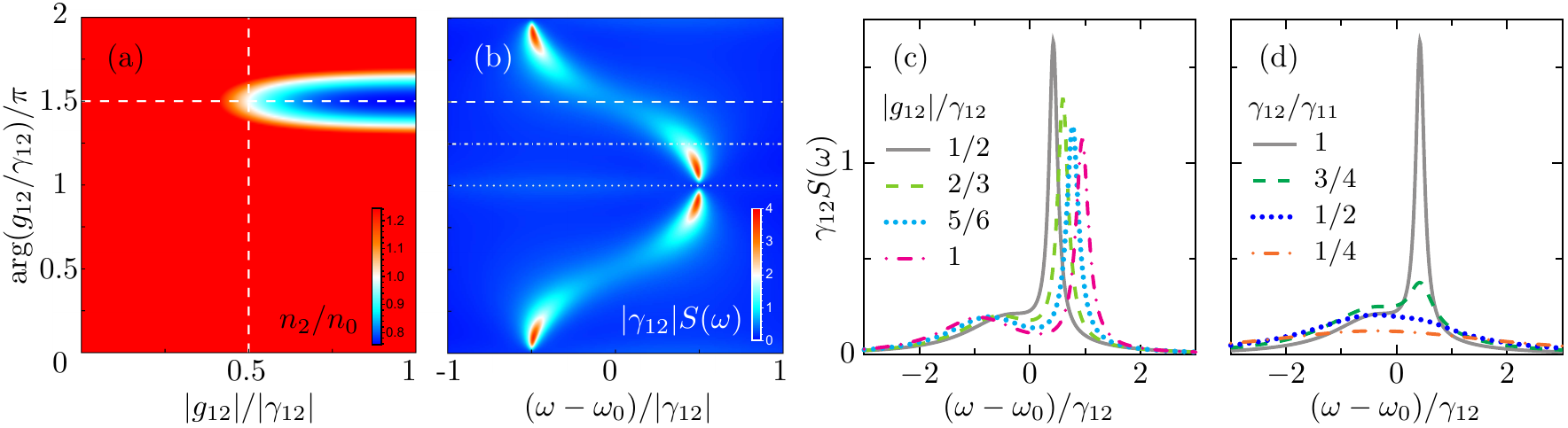}
\caption{(a) Steady-state population of QE-2, normalized to its
  population in isolation, as a function of the relative magnitude and
  phase of the coherent and dissipative couplings. White dashed lines
  plot Eqs.~\eqref{eq:conditions}. (b) Far-field spectrum $S(\omega)$
  as a function of $\mathrm{arg}(g_{12}/\gamma_{12})$ for $|g_{12}| =
  |\gamma_{12}|/2$ ($P_1 = P_2 = \gamma_{11} /100$ and $|\gamma_{12}|
  = \gamma_{11}$). The three horizontal lines indicate the
  configurations in Fig.~\ref{fig:fig3}. (c) $S(\omega)$ for
  increasing ratios of $|g_{12}| / \gamma_{11}$, with
  $\mathrm{arg}(g_{12}/\gamma_{12}) = 5 \pi /4$ and $\gamma_{12} =
  \gamma_{11}$. (d) $S(\omega)$ for decreasing
  $|\gamma_{12}|/\gamma_{11}$, with $\mathrm{arg}(g_{12}/\gamma_{12})
  = 5 \pi /4$, $|g_{12}| = |\gamma_{12}|/2$. The grey lines in (c) and
  (d) are evaluated along the white dot-dashed (middle) line in panel
  (b).} \label{fig:fig2}
\end{figure*}

Figure~\ref{fig:fig1}(b,c) analyze the deviation of SP-assisted
coupling from Eqs.~\eqref{eq:conditions} as a function of the QE
distance and natural frequency ($z=5$ nm). We observe that
both the magnitude ratio (b) and phase difference (c) undergo
oscillations around the chiral condition (white regions), which
become stronger as $x/\lambda_0$ and $\omega_0$ decrease. Both
maps are obtained from analytical expressions~\cite{supp}, and
demonstrate the possibility of varying the degree of chirality in
QE coupling exclusively through SPs. Red solid lines in
Figs.~\ref{fig:fig1}(d) and (e) plot cuts at $\omega_0 =
2.35~\mathrm{eV}$ in panels (b) and (c), respectively. Exact
numerical calculations for $\gamma_{\rm D}=0.03$ eV and increasing
$z$ are shown in dashed green (30 nm), dotted blue (60 nm),
dot-dashed orange (150 nm), and thin pink (300 nm) lines. The
deviations from Eqs.~\eqref{eq:conditions} are
larger than in the upper panels, as a result of the damping
experienced by SPs in their propagation. This demonstrates that
the tunability of the system increases further when the
contribution of free-propagating modes to the DGF becomes comparable to SPs.

%===========================================================================
%===========================================================================
%===========================================================================

Now we have proven the tunability of emitter-emitter interactions in
our EM set-up, we investigate its influence on quantum optical
properties. Figure~\ref{fig:fig2}(a) reveals the
hallmark of chirality in the QE populations. It shows the
steady-state population of QE-2, $n_2 = \langle \sigma_2^{\dagger}
\sigma_2 \rangle$, normalized to its population in isolation, $n_0
= P_2/(P_2+\gamma_{22})$, versus $|g_{12}|/|\gamma_{12}|$ and
$\mathrm{arg}(g_{12}/\gamma_{12})$. From now on, and unless stated
otherwise, we take $\gamma_{12}=|\gamma_{12}|=\gamma_{11}$. As anticipated,
when Eqs.~\eqref{eq:conditions} are fulfilled (intersection of the horizontal and vertical
long-dashed white lines) the backaction is eliminated (the output of
QE-2 drives QE-1 but the opposite is prevented) and one indeed finds $n_2=n_0$.
This is a manifestation of chiral coupling and is realizable in the
proposed emitter-emitter setup due to its nontrivial field configurations.
We note that $n_2=n_0$ along the white ring
in Fig.~\ref{fig:fig2}(a), which always includes the chiral point,
however its aspect ratio depends strongly on the system
parameters. Inside (outside) this ring, we find $n_2/n_0$ smaller
(larger) than unity.

We focus now on the parameter range given by the vertical white
dashed line in Fig.~\ref{fig:fig2}(a), and investigate the
properties of the coupled QEs when only the magnitude condition
for chirality is met. As shown in Fig.~\ref{fig:fig1}, this regime
is significantly more accessible than the chiral configuration, as
the geometric (QEs height and separation) and material (QE natural
frequency) constraints on the system are greatly relaxed. We
explore the system through the normalized power
spectrum~\cite{Lopez2016} $S(\omega) =\lim_{t
  \rightarrow \infty} \mathrm{Re} \{ \int_0^{\infty} \langle
\xi^{\dagger} (t) \xi (t+\tau) \rangle e^{\mathrm{i} \omega \tau}
\mathrm{d}\tau \} / (\pi n_\xi) $, which accounts for coherent
superposition of the photon emission by both QEs, with $\xi =
(\sigma_1 + \sigma_2)/\sqrt{2}$ and
$n_\xi=\langle\xi^{\dagger}\xi\rangle$. Figure~\ref{fig:fig2}(b)
plots $S(\omega)$ versus $\mathrm{arg}(g_{12}/\gamma_{12})$ for
$|g_{12}| = |\gamma_{12}|/2$. It reveals that, for certain phase
differences, an extremely narrow peak emerges that completely
dominates the spectral properties of the system. We term the
configuration yielding this sharp spectral feature as
\emph{quasi-chiral}, which smoothly evolves into the drastically
lower and broader single-peaked spectrum characteristic of the
\emph{reciprocal} [$\mathrm{arg}(g_{12}/\gamma_{12}) =
0,\pi,2\pi$] and \emph{chiral} [$\mathrm{arg}(g_{12}/\gamma_{12})
= \pi/2,3\pi/2$] regimes. Figures~\ref{fig:fig2}(c)~and~(d)
analyze the sensitivity of the quasi-chiral peak to
$|g_{12}|/|\gamma_{12}|$ and $|\gamma_{12}|/\gamma_{11}$,
respectively. The grey solid spectrum in both panels is a cut
along the white dot-dashed line in panel (b)
[$\mathrm{arg}(g_{12}/\gamma_{12}) = 5\pi/4$]. Upon increasing the
EM coupling strength, and thus away from the chiral condition, we
find a remarkable robustness in the spectrum, which still presents
a (slightly blue-shifted) prominent peak for
$|g_{12}|=|\gamma_{12}|$ [panel (c)]. Contrarily, by
decreasing the dissipative coupling $S(\omega)$ flattens, which
indicates the crucial role the $\beta$-factor plays in the
emergence of the quasi-chiral phenomenology [panel (d)].

\begin{figure}[!b]
 \includegraphics[width=0.95\linewidth]{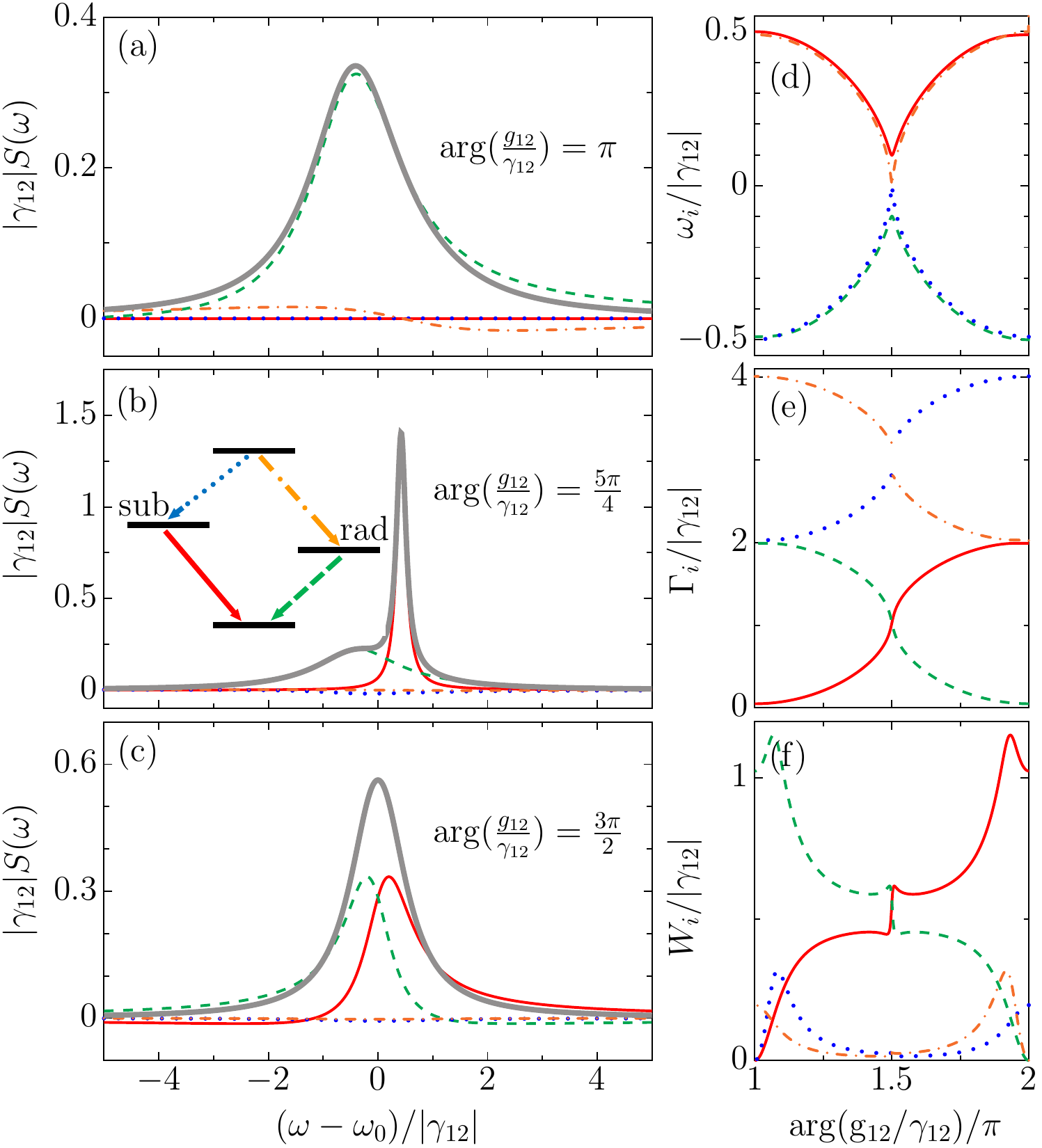}
 \caption{Left panels: $S(\omega)$ (thick gray lines) and its
constituents, $S_i(\omega)$ (thin colored lines), for the three
system configurations indicated by horizontal white lines in
Fig.~\ref{fig:fig2}(b): reciprocal (a), quasi-chiral (b), and
chiral (c). Right panels: spectral position (d), width (e) and
weight (f) of $S_i(\omega)$ as a function of $\mathrm{arg}
(g_{12}/\gamma_{12})$. In all panels, $P_1 = P_2 = \gamma_{11}
/100$, $|g_{12}| = |\gamma_{12}|/2$, and $\gamma_{12} \simeq
\gamma_{11}$. }
 \label{fig:fig3}
\end{figure}

We analyze in more detail the emission spectrum of the two coupled
QEs. For that purpose, we study the three different EM coupling
configurations marked by horizontal lines in
Fig.~\ref{fig:fig2}(b): reciprocal
[$\mathrm{arg}(g_{12}/\gamma_{12}) =\pi$], quasi-chiral [$5
  \pi /4$], and chiral [$3 \pi /2$]. The quasi-chiral condition is fulfilled
at the two shaded circles in Fig.~\ref{fig:fig1}(d-e),in which the
interaction between the QEs is highly directional, although not
fully unidirectional~\cite{supp}. The emission spectra for the
reciprocal, quasi-chiral and chiral regimes are shown as the solid
grey lines in Fig.~\ref{fig:fig3}(a), (b) and (c), respectively
(note the different vertical scales). The panels also display the
decomposition $S (\omega) = \sum_{i=1}^4 S_i
(\omega)$~\cite{delValle2010,supp}, where the contributions
$S_i(\omega)$ (non-solid color lines) arise from the four
transitions that can take place in the system. They are
illustrated by the diamond level scheme in the inset of
Fig.~\ref{fig:fig3}(b), with vertices at frequencies $\{0,
\omega_{\mathrm{rad}},\omega_{\mathrm{sub}}, 2\omega_0 \}$. Note
that $\omega_{\mathrm{rad}}$ and $\omega_{\mathrm{sub}}$ result
from the diagonalization of the Liouvillian~$\mathscr{L}$ and,
therefore, vary with the parameters. We observe that the
reciprocally coupled spectrum [panel (a)] is governed by a single
peak centered at $\omega_{\mathrm{rad}} \simeq \omega_0-|g_{12}| =
\omega_0-|\gamma_{12}|/2$ which originates from the transition
from the radiant to the ground state (green dashed line). Note
that $\mathrm{arg}(g_{12}/\gamma_{12}) =\pi$ in this case, which
yields $\omega_{\rm rad}<\omega_0$~\cite{supp}. The chiral system
[panel (c)] presents a single, broader maximum which emerges due
to the spectral overlapping of the emission of the two
single-excitation states
($\omega_{\mathrm{rad}}=\omega_{\mathrm{sub}}=\omega_0$). The two
QEs are only weakly coupled and $S(\omega)$ resembles the single
QE spectrum. Finally, we find that, surprisingly, the sharp peak
in the quasi-chiral emission [panel (b)] originates from the decay
of the subradiant state, with $\omega_{\mathrm{sub}} \simeq
\omega_0+0.4|\gamma_{12}|$. Note that a second, much lower and
broader maximum occurs at $\omega_{\mathrm{rad}} \simeq
\omega_0-0.4|\gamma_{12}|$.

To examine the causes of the evolution of the emission spectrum
with $\mathrm{arg}(g_{12}/\gamma_{12})$, we plot in the right
column of Fig.~\ref{fig:fig3} the resonant frequencies $\omega_i$
(d), the decay rates $\Gamma_i$ (e), and the weights $W_i$ (f) for
the four $S_i(\omega)$ contributions~\cite{delValle2010, supp}. We
can observe that by varying the phase difference between coherent
and dissipative couplings from the reciprocal to the chiral
configuration, the frequency and decay rate of the radiant and
subradiant states merge into the single QE values: $\omega_0$ and
$\gamma_{12}=\gamma_{11}$. In the evolution, the weight of the
subradiant contribution increases faster than its linewidth, which
gives rise to the formation, narrowing and blue-shifting of the
quasi-chiral spectral maximum shown in Fig.~\ref{fig:fig2}(b). In
particular, Fig.~\ref{fig:fig3}(e) evidences that the linewidth of
the sub-radiant contribution is much smaller than $\gamma_{12}$ at
$\mathrm{arg}(g_{12}/\gamma_{12}) =5\pi/4$, which gives rise to
the sharp spectral peak in the quasi-chiral spectrum.

\begin{figure}[t]
 \includegraphics[width=0.95\columnwidth]{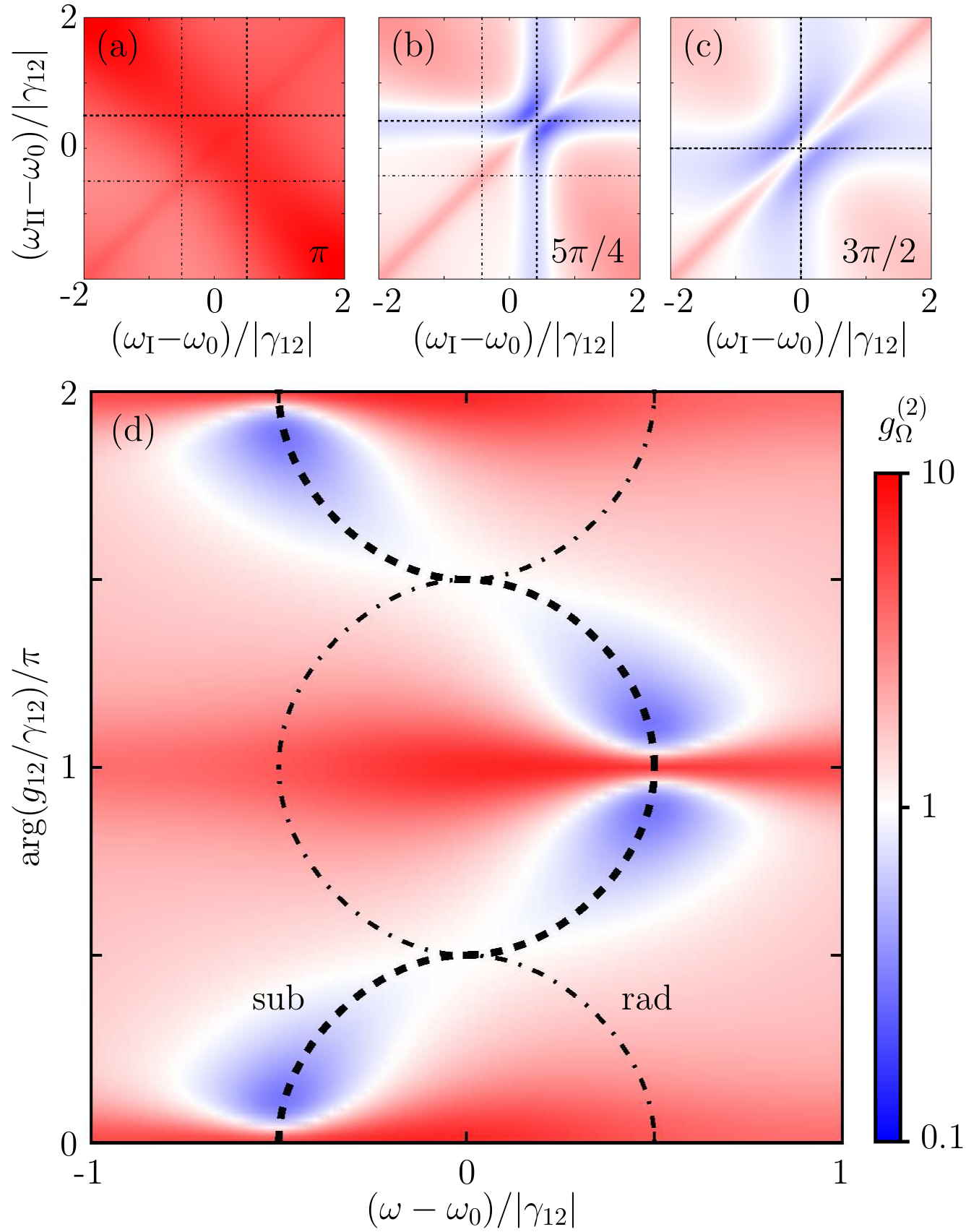}
 \caption{Two-photon spectra at zero time
delay, $g_\Omega^{(2)}$, as a function of the emission frequencies
$\omega_{\rm{I}}$ and $\omega_{\rm{II}}$. The three cases in
Fig.~\ref{fig:fig3} are shown: (a) reciprocal, (b) quasi-chiral,
and (c) chiral. The system is resonantly excited with a laser,
with filter size $\Omega = |\gamma_{12}|/5$. (d) Autocorrelation
function, $g_\Omega^{(2)}(\omega_{\rm{I}}=\omega_{\rm{II}}=\omega)$, versus
$\mathrm{arg}(g_{12}/\gamma_{12})$. In all panels, the thick dashed (thin dot-dashed)
lines track the (sub)radiant state.}
 \label{fig:fig4}
\end{figure}

%===========================================================================
%===========================================================================
%===========================================================================

In order to access the emission dynamics of the coupled QEs at a
deeper level, we investigate the two-photon correlations in
frequency, via the theory of two-photon
spectra~\cite{delValle2012}. Using the formalism of
Ref.~\cite{Gonzalez2013}, we compute the two-photon spectrum
$g_{\Omega}^{(2)} (\omega_{\mathrm{I}}, \omega_{\mathrm{II}})$ of our
system for photons detected at frequencies $\omega_{\rm{I, II}}$ (for more details, see Sec.~3 of Ref.~\cite{supp}).
We assume zero time delay and access the
correlations via detectors with spectral width $\Omega =
|\gamma_{12}|/5$. We plot in Fig.~\ref{fig:fig4}~(a, b, c) spectra
for the same $\mathrm{arg}(g_{12}/\gamma_{12})$ as in
Fig.~\ref{fig:fig3}~(a, b, c). In the three panels, a dashed thick
line marks the subradiant peak position and the dot-dashed thin
one marks the radiant frequency. The reciprocal system (a) broadly
demonstrates bunching [red regions, $g_{\Omega}^{(2)}>1$], with
little fine structure. However, upon evolving through the
quasi-chiral configuration (b) towards the chiral limit (c), a
remarkable butterfly structure appears: two bands of antibunching
[blue regions, $g_{\Omega}^{(2)}<1$] pierced by a diagonal line of
higher values of $g_{\Omega}^{(2)}$. This is the typical
manifestation of a single-photon emission, in this case at the
sub-radiant frequency~\cite{Gonzalez2013}. In panel (d), we map
the autocorrelation function $g_{\Omega}^{(2)}
(\omega_{\rm{I}}=\omega_{\rm{II}}=\omega)$ versus the phase
difference between coherent and dissipative coupling strengths.
The close resemblance to the one-photon spectrum of
Fig.~\ref{fig:fig2}(b) is immediately apparent. We conclude by
inspection that the minimum of
$g_{\Omega}^{(2)}$ overlaps almost exactly with the subradiant
peak position. Most notably, the global minimum in the correlation
function does not occur at the chiral conditions
but when the two QEs are quasi-chirally coupled, further
highlighting the singular optical properties that emerge in the
quasi-chiral regime.

%===========================================================================
%===========================================================================
%===========================================================================

To conclude, we have investigated the emergence of chirality in
the interactions between two quantum emitters. Exploring the
interplay between coherent and dissipative mechanisms, we find
naturally the non-reciprocal coupling configuration, which enables
us to identify the physical conditions for its occurrence. Through
analytical and numerical EM calculations, we have
shown that tunable chiral interactions can be realized in a
nanoscale platform consisting of two circularly-polarized emitters
held above a metal surface. Finally, we have unveiled the rich
quantum optical properties of the quasi-chiral regime, in which
the conditions for non-reciprocity are only partially met. We have
shown that the subradiant state of the system governs its one- and
two-photon spectra, which, remarkably, gives rise to both intense
emission and strong photon correlations within the same spectral
window. Our findings set the theoretical grounds and provide
guidance towards the development and optimization of quantum
optical functionalities associated with the fine tuning between
coherent and dissipative light-matter interactions, beyond the
highly stringent chiral regime.

This work has been funded by the Spanish MINECO under contract
FIS2015-64951-R (CLAQUE) and through the 'Mar\'ia de Maeztu' programme for
Units of Excellence in ${\rm R}\&{\rm D}$ (MDM-2014-0377). AIFD
acknowledges funding from the EU Seventh Framework Programme under
Grant Agreement FP7-PEOPLE-2013-CIG-630996.

\end{document}